\documentclass{ws-ijmpe}
\usepackage[super,compress]{cite}
\usepackage{hyperref}
\begin{document}

\markboth{Anthony W. Thomas}{Reflections on the Origin of the EMC Effect}

\catchline{}{}{}{}{}

\title{Reflections on the Origin of the EMC Effect}

\author{Anthony W. Thomas}

\address{CSSM and CoEPP, Department of Physics, The University of Adelaide\\
Adelaide, SA 5005,
Australia}



\maketitle


\begin{abstract}
In the 35 years since the European Muon Collaboration announced the astonishing result that the valence 
structure of a nucleus was very different from that of a free nucleon, many explanations have been 
suggested. The first of the two most promising explanations is based upon the different effects of the strong 
Lorentz scalar and vector mean fields known to exist in a nucleus on the internal structure of the nucleon-like clusters which occupy shell model states. The second links the effect to the modification of the structure of nucleons involved 
in short-range correlations, which are far off their mass shell. We explore some of the methods which 
have been proposed to give complementary information on this puzzle, especially the spin-dependent 
EMC effect  and the isovector EMC effect, both proposed by Clo\"et, Bentz and Thomas. It is shown that 
the predictions for the spin dependent EMC effect, in particular, differ substantially within the mean-field 
and short-range correlation approaches. Hence the measurement of the spin dependent EMC effect 
at Jefferson Lab should give us a deeper understanding of the origin of the EMC effect and, indeed, 
of the structure of atomic nuclei.
\end{abstract}


\ccode{PACS numbers: 24.85.+p, 25.30.Mr, 24.10.Jv, 11.80.Jy, 12.39.K}


\section{Dedication}

This manuscript is part of a volume intended to honor Professor Ernest Henley and his remarkable 
contributions to nuclear physics. I am especially pleased to be able to contribute as Ernest played a 
very significant role in my early development as a nuclear theorist. Since I was appointed, with Harold 
Fearing, to start the Theory Group at TRIUMF at the tender age of 25, Erich Vogt realized that while I 
knew rather a lot about pion-deuteron scattering and solving the Faddeev equations there were 
considerable gaps in my knowledge, especially when it came to experimental physics. So I was appointed 
as Secretary of the TRIUMF Experiment Evaluation Committee (EEC), commonly known as 
a PAC at other laboratories. In the late 70's and early 80's the EEC typically considered 10-20 new 
experimental proposals at each meeting (held twice per year), 
on topics ranging from nucleon-nucleon scattering to pion nucleus
scattering, rare muon decays, the measurement of Michel parameters, muonic atoms and $\mu$-SR. 

The physics being probed ranged from traditional nuclear physics, to strong interaction phenomena 
for which the meson factories were especially suited, to the search for physics beyond the Standard Model. 
Thus I was immersed at an impressionable age in a wonderful playground full of exciting new ideas and 
outstanding physicists, with experimental proposals from David Bugg, Ken Crowe and Wim van Oers amongst many others. The discussions in the closed sessions of the EEC provided a remarkable education 
as well as a unique opportunity to grow as a physicist. Amongst the 
experimenters on the EEC who provided their 
insights, I particularly remember Andy Bacher and Jules Deutsch as well as Alan Astbury a little later. But as 
a theorist the person from whom I learned the most was Ernest Henley. Time and again he provided deep 
new perspectives on the importance of an experimental proposal, explaining carefully to the rest of us the 
information that one might hope to extract and why it mattered in the broader context of subatomic 
physics. He was especially interested in symmetries and symmetry breaking as a way of winning new 
information about a system and, for example, he was an enthusiastic supporter of Wim van Oers' 
ambitious program of measurements of charge symmetry violation in the nucleon-nucleon system. Certainly 
a good part of the motivation for my later work in symmetry violation stemmed from this early exposure 
to Ernest. 

There were many later interactions with Ernest, although we never actually jointly authored a paper. He 
visited my group in Adelaide for an enjoyable month. At the time when the INT was being formed and 
Ernest was acting Director he invited Jerry Miller and me to organise the first workshop and I believe that 
the topics covered included the EMC effect. It is a pleasure to dedicate this manuscript as a tribute to Ernest.

\section{Introduction}

My introduction to the EMC effect~~\cite{Aubert:1983xm,Bodek:1983ec,Geesaman:1995yd,Malace:2014uea} 
may be of some interest. 
In 1982 I was a relatively young staff member in the 
Theory Division at CERN and by chance the office of the EP Division Leader, Erwin Gabathuler, was just 
down the corridor. One day Erwin walked into my office and after introducing himself proceeded to show me 
the puzzling data they had on nuclear structure functions. I had been chosen because I was one of the few 
theorists at CERN at the time who was not afraid to be associated with nuclear physics. On that day I had no 
idea what might be causing the effect but I was fascinated by it and still am! Incidentally, I have had another 
member of the EMC tell me later how wrong it was for Erwin to discuss unpublished data with me in that way. 
In my view this sort of honest discussion of challenging new results, 
as they arise, is a vital part of the practice
of physics and I wish there was more of it. The discussion that day stimulated the exciting work with Magda 
Ericson~\cite{Ericson:1983um} and Chris Llewellyn Smith~\cite{LlewellynSmith:1983vzz} on 
the important question of whether an enhancement of the pion field in a nucleus might 
be responsible. This in turn led to many further calculations and new experiments and even though it appears 
at present as though this is not the explanation we all learned a great deal.

While many exciting and imaginative ideas about the origin of the EMC effect were explored soon after its 
discovery~\cite{Nachtmann:1983py,Close:1983tn,Kondratyuk:1984qj,Celenza:1984ew,Bickerstaff:1985mp,Frankfurt:1985cv,Krzywicki:1985fd,Saito:1985ct,Gupta:1985mg,Dunne:1985ks,Hoodbhoy:1986fn,Efremov:1986mt,Bickerstaff:1989ch}, 
very few are still widely considered viable. We focus on two possibilities here, namely the change in internal structure generated by the mean scalar field in a nucleus and alternatively the potential effect on highly correlated nucleons of being far off-mass shell. At the start I must confess a bias, albeit one that I like to believe is primarily based in physics, rather than the fact that it is the approach on which I have expended a great deal of effort. 

Briefly, we have known for 40 years, based upon studies using dispersion theory, that the intermediate range attraction between two nucleons is a Lorentz scalar~\cite{Cottingham:1973wt}. The dominant short distance repulsion has a Lorentz vector character. Relativistic studies of atomic nuclei, whether based upon Quantum 
Hadrodynamics~\cite{Serot:1997xg}, relativistic Brueckner-Hartree-Fock~\cite{vanDalen:2004pn}, or any other variation, invariably conclude that these two very different contributions are both remarkably large, typically between 300-500 MeV. In many ways, the average nuclear binding of 8 MeV per nucleon is then the result of an accidental cancellation. The dynamical effect of scalar and vector potentials is very 
different~\cite{Guichon:1987jp,Thomas:2016bxx}, with the simplest implementation of the vector field simply shifting the energy scale, while the scalar potential modifies the masses and hence the dynamics of the confined quarks dramatically. It is the latter which, within the framework of QMC (the quark-meson coupling 
model~\cite{Guichon:1987jp,Guichon:1995ue,Saito:2005rv}), leads to important changes in the internal structure of a bound nucleon. In comparison with this  scalar field, with a strength up to half the mass of the nucleon itself, the degree to which one nucleon in a correlated pair, say with relative momentum 400 MeV/c, is off-shell is considerably smaller (around 80 MeV). This is the quantitative reason why I expect the dominant contribution to the EMC effect to come from  mean-field effects.

Over the past 30 years numerous applications of the QMC 
model~\cite{Saito:2005rv,Guichon:2008zz,Cloet:2015tha,Tsushima:1998qw}, both in its original form based upon the 
MIT bag model~\cite{Guichon:1987jp,Guichon:1995ue} and in its later re-incarnation starting with the covariant 
NJL model~\cite{Bentz:2001vc}, have shown that it can describe the EMC effect quantitatively across the periodic 
table~\cite{Thomas:1989vt,Saito:1992rm,Miller:2001tg,Mineo:2003vc,Cloet:2005rt,Cloet:2006bq,Cloet:2009qs}. The predictive capacity of the short-range correlation (SRC) idea is much less developed but it certainly is consistent with the observed dependence on A~\cite{Weinstein:2010rt,Hen:2016kwk}. What is clearly important is to find new, as yet unobserved phenomena, related to the EMC effect, for which these approaches make definite predictions. In this way one can hope to distinguish between them. We consider two examples here, the spin dependent EMC effect and the isovector EMC effect, both originally suggested by Clo\"et {\it et al.}.
\begin{figure}[th]
\centerline{\includegraphics[width=12cm]{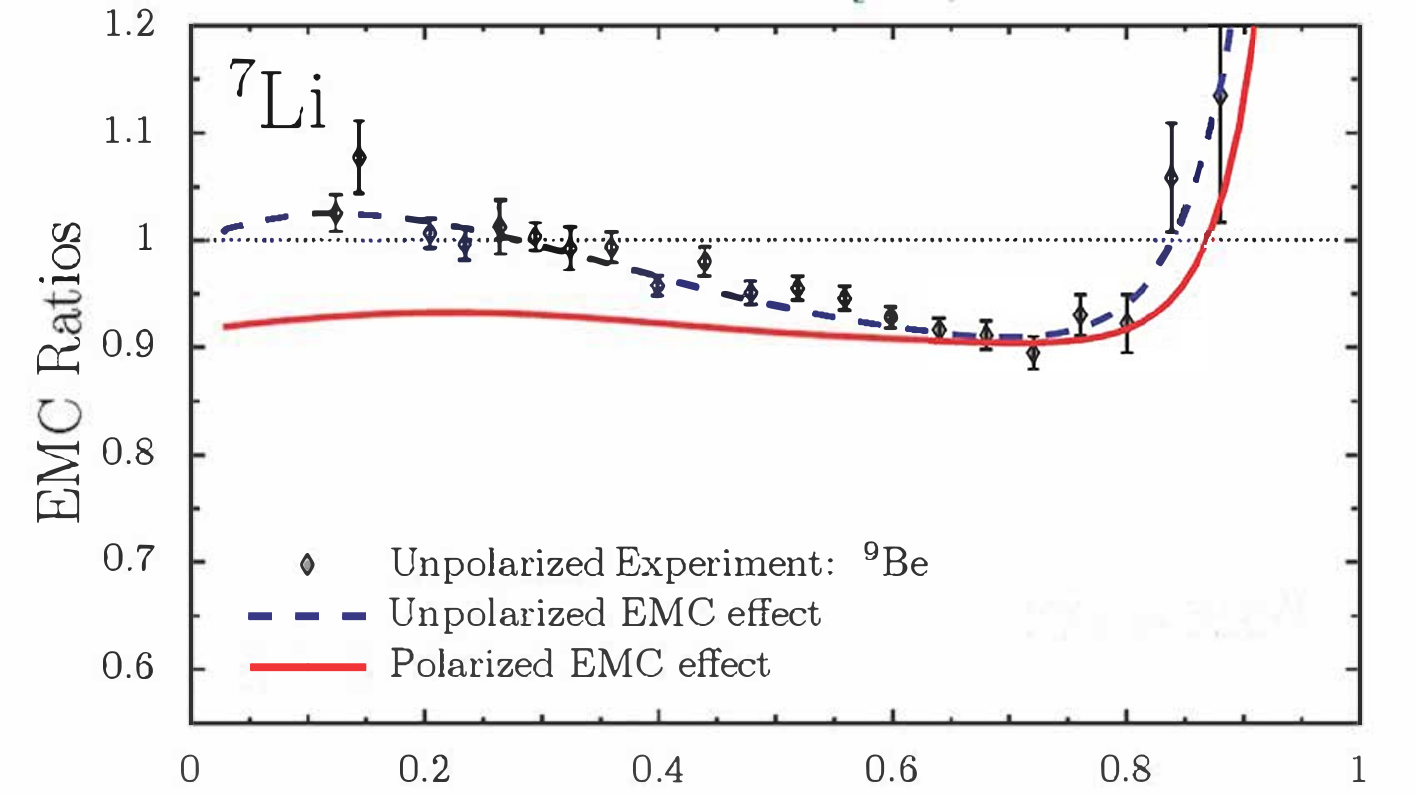}}
\caption{Predictions of the EMC effect in $^7$Li for the regular EMC effect (in comparison with data for $^9$Be) and for the spin dependent EMC effect. Note that the effect is shown with our understanding of the nuclear structure for $^7$Li used so that the proton polarization has been corrected to 100\%. The calculations were carried out by 
Clo\"et {\it et al.}~\cite{Cloet:2006bq} using the NJL model to describe the nucleon structure 
and are reported at a scale of 5 GeV$^2$.}
\label{fig:fig1}
\end{figure}

\section{The Spin EMC Effect}
The history of strong interaction physics has often demonstrated that the measurement of a spin dependent quantity can destroy a perfectly good theory. Alternatively, just occasionally, it may re-inforce one's faith in one. This line of thinking  led to the suggestion that one might gain important insight into the EMC effect by measuring the spin structure function of a selected nucleus. This is a difficult measurement because in an ideal case the spin is carried essentially by a single, unpaired valence nucleon and it is therefore an effect of order $1/A$. In order to have a chance of interpreting the result one really must have control of the nuclear structure so that the effective polarization of this valence nucleon is well known. Furthermore, because the spin structure function of the proton, $g_1^p$, is significantly larger than that of the neutron, the nucleus really must be one with a valence proton. However, if all of these conditions can be satisfied one can hope to compare the change in the spin dependent structure function of a bound nucleon against theoretical models.
\begin{figure}[th]
\centerline{\includegraphics[width=10cm]{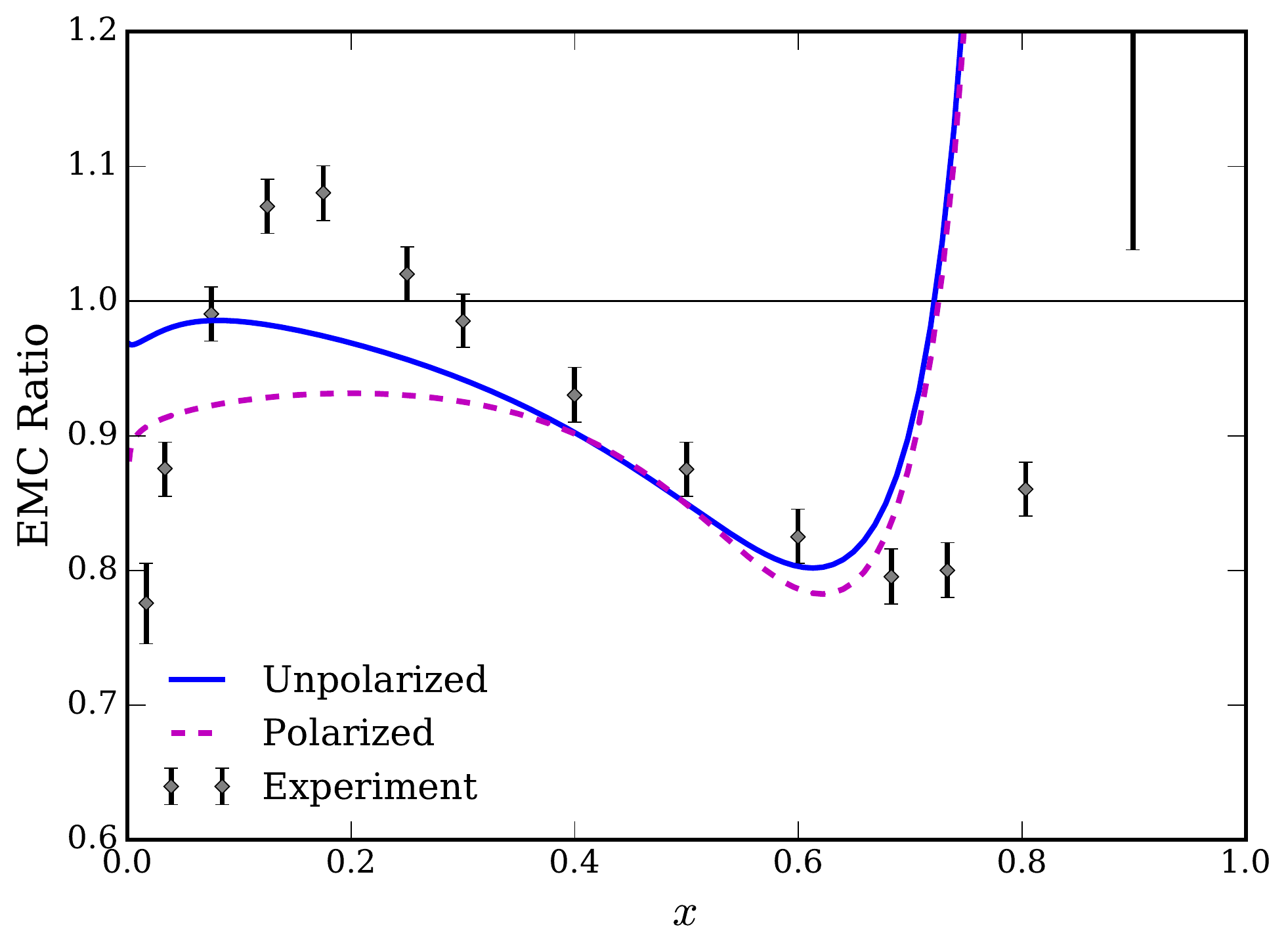}}
\caption{Comparison of the spin dependent and spin independent EMC effects for the idealised case of a proton in nuclear matter (at a scale of 10 GeV$^2$). The calculations were carried out within the QMC model as reported recently by 
Tronchin {\it et al.}~\cite{Tronchin:2018mvu}.}
\label{fig:tronchin}
\end{figure}

The first investigation of this effect by Clo\"et {\it et al.}~\cite{Cloet:2005rt,Cloet:2006bq} found the remarkable result that the polarized EMC effect, defined as the ratio of the structure function for a 100\% polarised bound proton (i.e., corrected for nuclear depolarization) to that of a free proton, was significantly larger than the unpolarized EMC effect. Even though Li is a little light for a mean field treatment, it is ideal for a first experiment and will be investigated at Jefferson Lab in the near future~\cite{bosted}. The prediction of Clo\"et {\it et al.}, shown in Fig.~\ref{fig:fig1}, is based upon the self-consistent modification of the structure of the bound proton in the NJL model and suggests that the spin EMC effect in Li should be very similar to the spin independent effect for the same nucleus.

Even though the QMC model was developed using the MIT bag model for the nucleon and this was indeed the basis for the first calculations of the EMC effect, there had been no calculations of the spin dependent EMC effect in that model until very recently. This lack was recently remedied~\cite{Tronchin:2018mvu} and Fig.~\ref{fig:tronchin} shows the resultant calculation of the EMC effects (spin independent and spin dependent) for a proton immersed in nuclear matter, using QMC. The key point is that both effects are clearly of the same size, so that in this approach, in which the internal structure of the bound nucleon is self-consistently modified in-medium, one does expect to find a sizeable polarized EMC effect. One aspect of the recent calculation which needs further work is that the spin saturation found earlier by Clo\"et {\it et al.}, using the NJL model was considerably larger than that found using the bag model. As this reduction will have important consequences for a number of current problems, such as neutrino-nucleus scattering and double $\beta$-decay, this needs to be resolved.

\subsection{Short-range correlations}
While the consequences for the spin dependent EMC effect of a modification of nucleon structure because of the large scalar mean field in-medium are clear, as illustrated above, the situation has been less clear for the case where SRC drive the EMC effect. We suggest that in the case of SRC, if they were indeed the sole source of the EMC effect, there should be little or no spin dependent EMC effect. This would make the up-coming measurement of the structure function of a polarized $^7$Li target a key test of this fundamental issue in modern nuclear physics.

It is well established through the important program of experiments at Jefferson Lab that the dominant source of high momentum nucleons in nuclei (i.e., well above the Fermi level) is 
tensor correlations~\cite{Subedi:2008zz,Arrington:2011xs}, involving a neutron-proton pair in the $^3S_1 - ^3D_1$ state. The angular momentum barrier means that the highest momentum nucleons in a correlated pair will be in $D$-wave. That is, they may meet with low relative momentum in a shell model configuration, where their relative  angular momentum is $S$-wave, but through SRC they will be scattered into a high relative momentum $D$-wave state by the tensor force. If this picture is correct, and it is certainly true that the $p^L$ behaviour of the relative wave function must support higher momentum in $D$-wave, then one can easily work out the effective polarization of a valence nucleon once it has scattered through a SRC into a high momentum state. 

Assuming the valence nucleon to be 100\% polarized and in $S$-wave with respect to the nucleon it strikes in the SRC, then the z-component of total angular momentum will be 50\% $J_z=0$ and 50\% $J_z=+1$. After the short-distance tensor interaction $J_z$ is preserved. However, the Clebsch-Gordan coefficients  for orbital angular momentum $L=2$ overwhelmingly favour the spin down configurations. This significantly depolarizes the correlated struck proton which is a fair way off-mass-shell because of the high momentum carried away by its partner. In fact, its polarization will be of order -10 to -15\% instead of +100\%. Thus the structure function of this weakly polarized proton cannot exhibit a strong spin dependence.

To summarize, SRC will depolarize the valence proton so that, if the EMC-type modification of structure functions does arise only through the change of the structure function of a high momentum, far off-mass-shell nucleon in a correlated pair, there can be very little EMC effect on the nuclear spin structure function. Once one divides the measured nuclear structure function by the effective nucleon polarization one expects little nuclear modification of the spin structure function and hence no significant spin-dependent EMC effect in this model.

\section{Isovector EMC Effect}
\begin{figure}[th]
\centerline{\includegraphics[width=12cm]{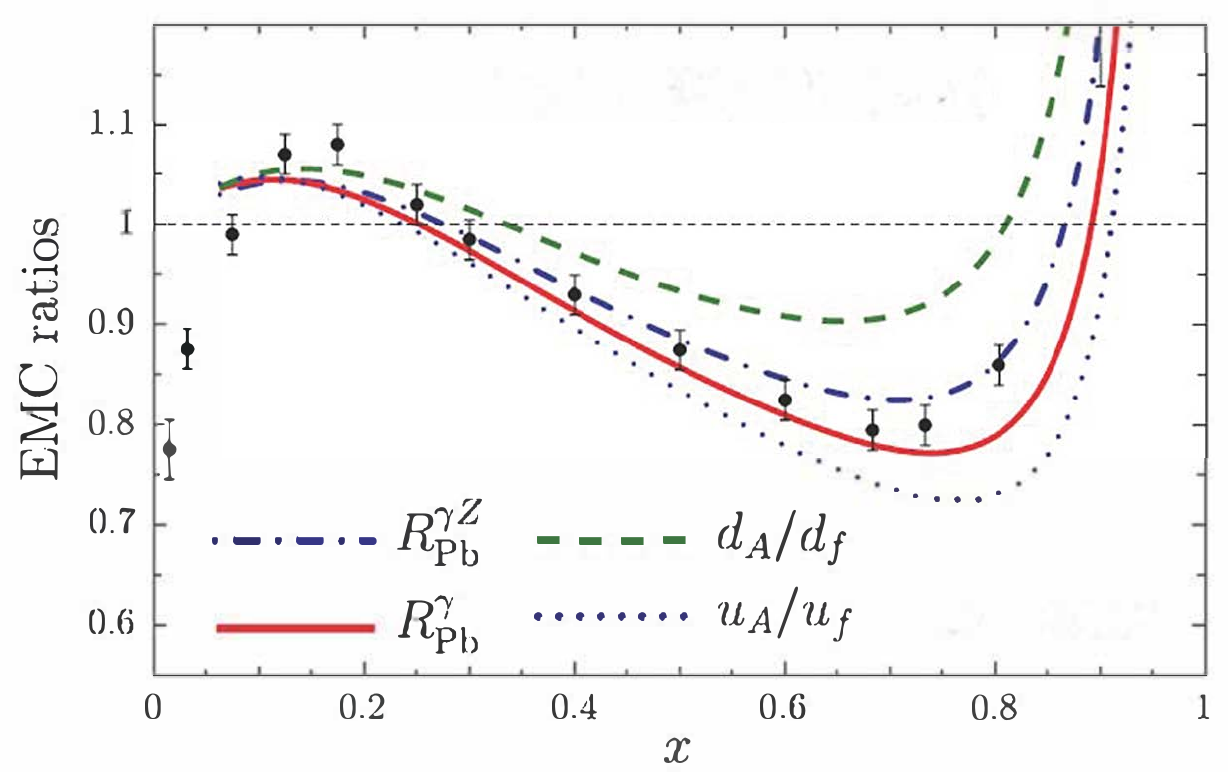}}
\caption{Illustration of the effects of the isovector nuclear force on structure functions and PDFs in nuclear matter with 
an $N/Z$ ratio corresponding to Pb (at a scale of 5 GeV$^2$), as calculated in Refs.~\cite{Cloet:2009qs,Cloet:2012td}. While the difference between the electromagnetic and parity violating structure functions should be measureable in a precise experiment, a direct determination of the separate $u$ and $d$ distributions would be definitive.}
\label{fig:fig3}
\end{figure}
Within a mean field picture of the EMC effect, the Lorentz vector potentials enter the calculation of structure functions in a relatively simple way. This led Clo\"et {\it et al.}~\cite{Cloet:2009qs,Cloet:2012td}
 to realize that the existence of an isovector mean field in a nucleus with N not equal to Z, which is well understood in nuclear structure physics, would shift the $d$ and $u$-quark PDFs in opposite ways, with the every one of the former, feeling repulsion from the excess of $d$-quarks in the nucleus, {\em regardless of whether they were in a proton or a neutron}. Similarly, all of the $u$-quarks must fell additional attraction {\em regardless of which nucleon they call home}. Thus, while this difference in energies is completely consistent with isospin symmetry, to a naive observer it looks remarkably like an increase in the well-known charge symmetry violation of the PDFs arising from the fact that $m_d>m_u$~\cite{Londergan:2009kj,Rodionov:1994cg}
. It naturally leads to a transfer of light-cone momentum from all of 
the $u$-quarks in a nucleus with $N\neq Z$ to the $d$-quarks. This observation can account for a significant fraction of the NuTeV anomaly~\cite{Cloet:2009qs,Bentz:2009yy}.

The rather significant effects of this shift of momentum are illustrated in Fig.~\ref{fig:fig3}. The definitive measurement would involve a separate determination of the total $u$ and the total $d$-quark distributions in a nucleus like Pb, where as shown the difference in the respective EMC ratios is as large as 20\%. Charged current weak interactions would be the ideal tool for studying this phenomenon. A measurement of the somewhat smaller difference between the parity-violating $\gamma-Z$ interference structure function and the normal electromagnetic structure function would also be extremely valuable. The latter is certainly feasible at Jefferson Lab following its 12 GeV upgrade.

At present I find the discussion of this phenomenon with the SRC approach confusing. Certainly the discussion focusses on the fraction of momentum carried by protons and neutrons, whereas in the mean field approach the shift of momentum from $u$ to $d$-quarks is independent of which nucleon contains the quark in question.

\section{Concluding Remarks}
The EMC effect is fundamental to our understanding of nuclear structure. Nuclear theory needs to provide a clear understanding of the results found, no matter what probe is used to investigate it. The theory used should also provide a quantitative description of nuclear structure. It is therefore gratifying that the QMC approach, in which the structure of a bound nucleon is self-consistently modified by the scalar fields generated by the other nucleons, which had early success in application to the EMC effect, has since been used to generate very realistic energy 
density functionals~\cite{Guichon:2006er,Guichon:2004xg}. Indeed, the energy density functional (EDF) derived from the QMC model has been shown to reproduce the binding energies of nuclei across the periodic table at the 0.3\% level, with superheavies even better at 0.1\%~\cite{Stone:2016qmi,Stone:2017oqt,Guichon:2018uew}. Quite reasonable results have also been obtained for charge radii, deformations and two-nucleon and alpha particle separation energies. It is particularly interesting that, while the level of accuracy is not quite as good as the best phenomenological EDFs, the number of parameters in the QMC approach is less than half the number typically used there. 

We have reviewed the predictions of the mean field approach to the EMC effect for new phenomena, such as the polarized  or spin dependent EMC effect and found that it is expected to be at least as large as the unpolarized EMC effect on a given nucleus. As our examination of the proposed explanation in terms of SRC suggest that this will produce little or no spin dependent EMC effect this is a critical test of the underlying physics. It makes the upcoming experimental test at Jefferson Lab particularly important.

A second important test of the physics of the EMC effect involves the so-called isovector EMC effect. In the mean field approach this involves a shift of momentum from every $u$-quark to every $d$-quark, regardless of whether it is in a proton or a neutron. Parity violating electron scattering and charged current weak interactions with nuclei are the ideal tools to investigate the predictions in this area and once again we look to Jefferson Lab for leadership in those investigations. 

After 35 years there are still many challenges associated with understanding the origin of the EMC effect. Nevertheless, the substantial theoretical and experimental progress in recent years gives us confidence that we can resolve these challenges before much longer. Our understanding of the structure of atomic nuclei will be much deeper as a result.

\section*{Acknowledgements}

This work was supported by the University of Adelaide and by the Australian Research Council through the 
ARC Centre of Excellence for Particle Physics at the Terascale (CE110001104) and Discovery Projects 
DP150103101 and DP180100497.

\end{document}